\newcommand{\bea}{\begin{eqnarray}}
\newcommand{\eea}{\end{eqnarray}}
\newcommand{\be}{\begin{equation}}
\newcommand{\ee}{\end{equation}}
\newcommand{\ben}{\begin{enumerate}}
\newcommand{\een}{\end{enumerate}}
\newcommand{\bi}{\begin{itemize}}
\newcommand{\ei}{\end{itemize}}
\newcommand{\bmi}[1]{\begin{minipage}{#1 cm}}
\newcommand{\emi}{\end{minipage}}

\newcommand{\vectii}[2]{\rund{\begin{array}{c} #1 \\ #2 \end{array} }}

\def\elabel#1{\label{eq:#1}}

\def\eck#1{\left\lbrack #1 \right\rbrack}

\def\rund#1{\left( #1 \right)}
\def\abs#1{\left\vert #1 \right\vert}

\def\ave#1{\left\langle #1 \right\rangle}

\def\Re{{\cal R}\hbox{e}}
\def\Im{{\cal I}\hbox{m}}

\def\D{{\cal D}}

\def\P{{\cal P}}

\def\d{{\rm d}}

\def\vp{\varphi}
\def\vt{{\vartheta}}

\def\Real{{\rm I\mathchoice{\kern-0.70mm}{\kern-0.70mm}{\kern-0.65mm}%
  {\kern-0.50mm}R}}
\def\C{\rm C\kern-.42em\vrule width.03em height.58em depth-.02em
       \kern.4em}
\font \bolditalics = cmmib10
\def\bx#1{\leavevmode\thinspace\hbox{\vrule\vtop{\vbox{\hrule\kern1pt
        \hbox{\vphantom{\tt/}\thinspace{\bf#1}\thinspace}}
      \kern1pt\hrule}\vrule}\thinspace}

\def \vc #1{{\textfont1=\bolditalics \hbox{$\bf#1$}}}
{\catcode`\@=11
\gdef\SchlangeUnter#1#2{\lower2pt\vbox{\baselineskip 0pt \lineskip0pt
  \ialign{$\m@th#1\hfil##\hfil$\crcr#2\crcr\sim\crcr}}}
}

\def\ueber#1#2{{\setbox0=\hbox{$#1$}%
  \setbox1=\hbox to\wd0{\hss$\scriptscriptstyle #2$\hss}%
  \offinterlineskip
  \vbox{\box1\kern0.4mm\box0}}{}}

\def\bx#1{\leavevmode\thinspace\hbox{\vrule\vtop{\vbox{\hrule\kern1pt
        \hbox{\vphantom{\tt/}\thinspace{\bf#1}\thinspace}}
      \kern1pt\hrule}\vrule}\thinspace}

{\catcode`\@=11
\gdef\SchlangeUnter#1#2{\lower2pt\vbox{\baselineskip 0pt \lineskip0pt
  \ialign{$\m@th#1\hfil##\hfil$\crcr#2\crcr\sim\crcr}}}
}

\documentclass{aa}
\usepackage{graphicx}
\begin{document}
   \title{B-modes in cosmic shear from source redshift clustering}

   \author{Peter Schneider
          \inst{1,2}
          \and
          Ludovic van Waerbeke\inst{3,4}
	  \and
	  Yannick Mellier\inst{3,5}
          }

   \offprints{P. Schneider}

   \institute{Institut f. Astrophysik u. Extr. Forschung, Universit\"at Bonn,
              Auf dem H\"ugel 71, D-53121 Bonn, Germany\\
              \email{peter@astro.uni-bonn.de}
         \and
 	 	Max-Planck-Institut f. Astrophysik, Postfach 1317,
              D-85741 Garching, Germany
	\and
               Institute d'Astrophysique de Paris, 98 bis, boulevard
              Arago, F-75014 Paris, France
         \and
		Canadian Institute for Theoretical Astrophysics, 60 St
              Georges Str., Toronto, M5S 3H8 Ontario, Canada
	\and
		Observatoire de Paris, DEMIRM/LERMA, 61 avenue de 
               l'Observatoire, F-75014 Paris, France
             }

   \date{Received ; accepted }

   \abstract{Weak gravitational lensing by the large scale
	structure can be used to probe the dark matter distribution in
	the Universe directly and thus to probe cosmological
	models. The recent detection of cosmic shear by several groups
	has demonstrated the feasibility of this new mode of
	observational cosmology. In the currently most extensive
	analysis of cosmic shear, it was found that the shear field
	contains unexpected modes, so-called B-modes, which are
	thought to be unaccountable for by lensing. B-modes can in
	principle be generated by an intrinsic alignment of galaxies
	from which the shear is measured, or may signify some
	remaining systematics in the data reduction and analysis. In
	this paper we show that B-modes in fact {\it are produced} by
	lensing itself. The effect comes about through the clustering
	of source galaxies, which in particular implies an angular
	separation-dependent clustering in redshift. After presenting
	the theory of the decomposition of a general shear field
	into E- and B-modes, we calculate their respective power
	spectra and correlation functions for a clustered source
	distribution. Numerical and analytical estimates of the
	relative strength of these two modes show that the resulting
	B-mode is very small on angular scales larger than a few
	arcminutes, but its relative contribution rises quickly
	towards smaller angular scales, with comparable power in both
	modes at a few arcseconds. The relevance of this effect with
	regard to the current cosmic shear surveys is discussed; it
	can not account for the apparent detection of a B-mode
	contribution on large angular scales in the cosmic shear
	analysis of van Waerbeke et al.\ (2002).
     }

   \maketitle
%

\section{Introduction}
Gravitational lensing by the large-scale structure (LSS) leads to the
distortion of the images of distant galaxies, owing to the tidal
gravitational field of the matter inhomogeneities. Following very
early work on the study of light propagation in an inhomogeneous
universe (e.g., Gunn 1967; Kantowski 1969), Blandford et al. (1991),
Miralda-Escude (1991) and Kaiser (1992) have pointed out that the
observation of this `cosmic shear' effect immediately yields
information about the statistical properties of the LSS and, thus, on
cosmology. Non-linear evolution of the matter spectrum was taken into
account in later analytical (e.g., Jain \& Seljak 1997; Bernardeau et
al. 1997, Kaiser 1998; Schneider et al. 1998, hereafter SvWJK) and
numerical (e.g., van Waerbeke et al. 1999; Jain et al. 2000, White \&
Hu 2000) studies; see Mellier (1999) and Bartelmann \& Schneider
(2001; hereafter BS01) for recent reviews.

It was only in 2000 when four teams nearly simultaneously and
independently announced the first detections of cosmic shear from
wide-field imaging data (Bacon et al. 2000; Kaiser et al. 2000; van
Waerbeke et al. 2000; Wittman et al. 2000). The detections reported in
these papers (and in Maoli et al. 2001, using the VLT, and Rhodes et
al. 2001, using HST images obtained with the WFPC2 camera) concerned
various two-point statistics, like the shear dispersion in an
aperture, or the shear correlation function. In van Waerbeke et
al. (2001), the aforementioned statistics, as well as the aperture
mass statistics (SvWJK), were inferred from the effective 6.5 square
degrees of high-quality imaging data. Very recently, H\"ammerle et al.\
(2002) reported on a cosmic shear detection using HST parallel images
taken with the STIS instrument on an effective angular scale of $\sim
30''$. 

The shear field, originating from the inhomogeneous matter
distribution, is a two-dimensional quantity, whereas the projected
density field of the matter is a scalar field. The relation between
the shear $\gamma(\vc\theta)=\gamma_1(\vc\theta)+{\rm
i}\gamma_2(\vc\theta)$ and the projected matter density
$\kappa(\vc\theta)$ is 
\be
\gamma(\vc\theta)={1\over\pi}\int_{\Real^2}\d^2\theta'\;
\D(\vc\theta-\vc\theta')\kappa(\vc\theta')\;,
\label{eq:1.1}
\ee
with the kernel
\be
\D(\vc\theta)={\theta_2^2-\theta_1^2-2{\rm i}\theta_1\theta_2
\over \abs{\vc\theta}^4}\;;
\label{eq:1.2}
\ee
here, $\kappa$ is the dimensionless surface mass density, i.e., the
physical surface mass density divided by the `critical' surface mass
density, as usual in gravitational lensing; we follow the notation of
BS01 in this paper. Since the two shear components originate from a
single scalar field, they are related to each other; in particular,
their partial derivatives should satisfy compatability relations, as
we shall discuss in Sect.\ts 2 below. In analogy with the polarization
of the CMB, a shear field satisfying these compatability relations is
called an E-mode shear field.

Pen et al. (2002) pointed out that the cosmic shear data of van
Waerbeke et al. (2001) contains not only an E-mode, but also a
statistically significant B-mode contribution in addition. Such
B-modes can be generated by effects unrelated to gravitational
lensing, such as intrinsic alignment of galaxies (e.g., Heavens et
al.\ 2000, Crittenden et al.\ 2001a; Croft \& Metzler 2000; Catelan et
al.\ 2000) or remaining systematics in the data reduction and
analysis.

In this paper we show that a B-mode contribution to the cosmic shear
is obtained by lensing itself. A B-mode is generated owing to the
clustering properties of the faint galaxies from which the shear is
measured. This spatial clustering implies an angular
separation-dependent clustering in redshift, which is the origin not
only of the B-mode of the shear, but also of an additional E-mode
contribution. 

The paper is organized as follows: in Sect.\ts 2 we provide a tutorial
description of the E/B-mode decomposition of a shear field. Most of
the results there were derived before in Crittenden et al.\ (2001b,
hereafter C01), but we formulate them in standard lensing notation,
which will be needed for the later investigation. 
The calculation of two-point cosmic shear statistics
in the presence of source clustering is presented in Sect.\ts 3 where
it is shown that this clustering produces a B-mode. Numerical and
analytical estimates of the amplitude of this B-mode are provided in
Sect.\ts 4 and discussed in Sect.\ts 5.

\section{E/B-mode decomposition of a shear field}
In this section we provide the basic relations for the decomposition
of the shear field into E- and B-modes. Most of these relations
have been obtained in C01; we shall write them here in standard
lensing notation.

\subsection{Motivation}
If the shear field is obtained from a projected surface mass density
$\kappa$ as in Eq.\ts(\ref{eq:1.1}), then the gradient of the density
field $\kappa$ is related to the first spatial derivatives of the
shear components in the following way (Kaiser 1995):
\be
\nabla\kappa=\vectii{\gamma_{1,1}+\gamma_{2,2}}{\gamma_{2,1}-\gamma_{1,2}}
\equiv \vc u
\elabel{N1}
\ee
The vector field $\vc u$ can be obtained from observations, e.g. in
weak lensing cluster mass reconstructions, by obtaining a smoothed
version of the shear field and then differentiating this
numerically. Owing to noise, the resulting (`observed') field $\vc u$
will in general not be a gradient field. The non-gradient part of $\vc
u$ is then a readily identifiable noise component and can be filtered
out in the mass reconstruction. Seitz \& Schneider (1996) provided a
scheme for this noise filtering (see also Seitz \& Schneider 2001 for
a simpler though equivalent method), which was shown by Lombardi \&
Bertin (1998) to be an optimal reconstruction method.

If the shear field cannot be ascribed to a single geometrically thin
gravitational lens, the non-gradient part of $\vc u$ is not
necessarily due to noise. For example, if the galaxies have intrinsic
alignments, this may induce a curl-part of $\vc u$. To project out the
gradient and curl part of $\vc u$, we take a further derivative 
of $\vc u$, and define
\be
\nabla^2\kappa^{\rm E}=\nabla\cdot\vc u \quad; \quad
\nabla^2\kappa^{\rm B}=\nabla\times\vc u\equiv
u_{2,1}-u_{1,2}\;.
\ee
Through these relations, $\kappa^{\rm E}$ and $\kappa^{\rm B}$ are not
uniquely defined on a finite data field; as discussed in Seitz \&
Schneider (1996), a further condition is needed to specify the two
modes uniquely. However, we shall not be concerned here with
finite-field effects.

An alternative way to define $\kappa^{\rm E}$ and $\kappa^{\rm B}$ is
through the Kaiser \& Squires (1993) mass-reconstruction relation
\be
\kappa^{\rm E}(\vc\theta)+{\rm i}\kappa^{\rm B}(\vc\theta)
={1\over\pi}\int_{\Real^2}\d^2 \theta'
\D^*(\vc\theta-\vc\theta')\,\gamma(\vc\theta')\;,
\label{eq:2.3}
\ee
which formally requires data on an infinite field; here, $\D^*$
denotes the complex-conjugate of the complex kernel (\ref{eq:1.2}). If
$\gamma$ is of the form (\ref{eq:1.1}) with a real field $\kappa$,
then the result from (\ref{eq:2.3}) will be real, $\kappa^{\rm
E}=\kappa$, $\kappa^{\rm B}=0$. In applications of the KS-formula
(\ref{eq:2.3}) to observational data, where the recovered shear field
necessarily is noisy, one usually takes the real part of the integral
to obtain the projected mass density field. For a general shear field,
the result from (\ref{eq:2.3}) will be complex, with the real part
yielding the E-mode, and the imaginary part corresponding to the
B-mode.

To simplify notation and calculations, it is convenient to express
two-component quantities in terms of complex numbers. We define the E-
and B-mode potentials $\psi^{\rm E}$ and $\psi^{\rm B}$ by
\be
\nabla^2\psi^{\rm E,B}=2\kappa^{\rm E,B}\;,
\ee
and combine the two modes into the complex fields
\be
\kappa=\kappa^{\rm E}+{\rm i}\kappa^{\rm B}\quad ,
\quad  \psi=\psi^{\rm E}+{\rm i}\psi^{\rm B}\;.
\ee
The complex shear $\gamma=\gamma_1+{\rm i}\gamma_2$ is obtained from
the potential $\psi$ by $\gamma={\rm D}\psi$, where the differential
operator ${\rm D}=(\partial_{11}-\partial_{22})/2+{\rm
i}\partial_{12}$; hence,
\[
\gamma=\eck{{1\over 2}\rund{\psi^{\rm E}_{,11}-\psi^{\rm E}_{,22}}
-\psi^{\rm B}_{,12}} +{\rm i}\eck{\psi^{\rm E}_{,12}
+{1\over 2}\rund{\psi^{\rm B}_{,11}-\psi^{\rm B}_{,22}}} \;.
\]
Inserting this into (\ref{eq:N1}) yields 
\[
\vc u=\vectii{\kappa^{\rm E}_{,1}-\kappa^{\rm B}_{,2}}
{\kappa^{\rm E}_{,2}+\kappa^{\rm B}_{,1}} \;.
\]
Indeed, the shear field can be decomposed into E/B-modes,
$\gamma=\gamma^{\rm E}+{\rm i}\gamma^{\rm B}$, with
\[
\gamma^{\rm E}={\D\over \pi} * \Re\eck{{\D^*\over \pi} * \gamma}\;,
\]
\[
\gamma^{\rm B}={\D\over \pi} * \Im\eck{{\D^*\over \pi} * \gamma}\;,
\]
where the operator $\D$ is defined in Eqs.\ts(\ref{eq:1.1}),
(\ref{eq:1.2}), and `*' denotes complex conjugation. Thus, the two
components can be obtained from the shear field by filtering, except
for an additive constant.

\subsection{Shear correlation functions and power spectra}
The discussion above dealt with the shear field itself. In the
application to cosmic shear, one usually does not investigate the
shear of a $\kappa$-field itself, but its statistical properties. In
this paper we shall concentrate solely on two-point statistical
measures of the cosmic shear, and their decomposition into E- and
B-modes. 

Owing to statistical homogeneity and isotropy of the Universe,
$\kappa^{\rm E,B}(\vc\theta)$ are homogeneous and isotropic random
fields. Hence, in terms of their Fourier transforms
\be
\hat\kappa^{\rm E,B}(\vc\ell)=\int\d^2\theta\,{\rm e}^{{\rm
i}\vc\ell\cdot\vc\theta}\,\kappa^{\rm E,B}(\vc\theta)\;,
\ee
one defines the two power spectra $P_{\rm E}$, $P_{\rm B}$, and the
cross power spectrum $P_{\rm EB}$ by
\bea
\ave{\hat\kappa^{\rm E}(\vc\ell)\hat\kappa^{\rm E*}(\vc\ell')}
&=&(2\pi)^2\,\delta_{\rm D}(\vc\ell-\vc\ell')\,P_{\rm E}(\ell)\;,
\nonumber \\
\ave{\hat\kappa^{\rm B}(\vc\ell)\hat\kappa^{\rm B*}(\vc\ell')}
&=&(2\pi)^2\,\delta_{\rm D}(\vc\ell-\vc\ell')\,P_{\rm B}(\ell)\;,
\\
\ave{\hat\kappa^{\rm E}(\vc\ell)\hat\kappa^{\rm B*}(\vc\ell')}
&=&(2\pi)^2\,\delta_{\rm D}(\vc\ell-\vc\ell')\,P_{\rm EB}(\ell)\;,
\nonumber
\eea
where $\delta_{\rm D}$ denotes Dirac's delta distribution.
In terms of the complex field $\kappa$, we then have
\bea
\ave{\hat\kappa(\vc\ell)\hat\kappa^{*}(\vc\ell')}
&=&(2\pi)^2\,\delta_{\rm D}(\vc\ell-\vc\ell')\eck{P_{\rm
E}(\ell)+P_{\rm B}(\ell)}\;,
\nonumber \\
\ave{\hat\kappa(\vc\ell)\hat\kappa(\vc\ell')}
&=&(2\pi)^2\,\delta_{\rm D}(\vc\ell+\vc\ell')
\\
&\times& \eck{P_{\rm
E}(\ell)-P_{\rm B}(\ell)+2{\rm i}P_{\rm EB}(\ell)}\;.\nonumber
\eea
The Fourier transform $\hat\gamma(\vc\ell)$ of the shear is related to
$\hat\kappa(\vc\ell)$ through
\be
\hat\gamma(\vc\ell)=\rund{\ell_1^2-\ell_2^2+2{\rm i}\ell_1\ell_2
\over \abs{\vc\ell}^2}\hat\kappa(\vc\ell)
={\rm e}^{2{\rm i}\beta}\hat\kappa(\vc\ell)\;,
\elabel{N10}
\ee
where $\beta$ is the polar angle of $\vc\ell$. The correlators of the
shear then become
\bea
\ave{\hat\gamma(\vc\ell)\hat\gamma^*(\vc\ell')}
&=&(2\pi)^2\,\delta_{\rm D}(\vc\ell-\vc\ell')\eck{P_{\rm
E}(\ell)+P_{\rm B}(\ell)}\;,
\nonumber \\
\ave{\hat\gamma(\vc\ell)\hat\gamma(\vc\ell')}
&=&(2\pi)^2\,\delta_{\rm D}(\vc\ell+\vc\ell')\,{\rm e}^{4{\rm i}\beta}
\\
&\times& \eck{P_{\rm E}(\ell)-P_{\rm B}(\ell)+2{\rm i}P_{\rm
EB}(\ell)}\;. \nonumber
\eea
Next we define the correlation functions of the shear. This is done by
considering pairs of positions $\vc\vt$ and $\vc\theta+\vc\vt$, and
defining the tangential and cross-component of the shear at position
$\vc\vt$ for this pair as
\be
\gamma_{\rm t}=-\Re\rund{\gamma\,{\rm e}^{-2{\rm i}\vp}} \;,\quad
\gamma_{\times}=-\Im\rund{\gamma\,{\rm e}^{-2{\rm i}\vp}} \;,
\ee
respectively, where $\vp$ is the polar angle of the separation vector
$\vc \theta$. Then, the shear correlation functions are defined as  
\bea
\xi_+(\theta)&=&\ave{\gamma_{\rm t}\gamma_{\rm t}} +\ave{\gamma_\times
\gamma_\times}(\theta)\;,\nonumber\\
\xi_-(\theta)&=&\ave{\gamma_{\rm t}\gamma_{\rm t}} -\ave{\gamma_\times
\gamma_\times}(\theta)\;,
\elabel{corr}\\
\xi_\times(\theta)&=&\ave{\gamma_{\rm t}\gamma_{\times}}(\theta)
\;.\nonumber 
\eea
The shear correlation functions are most easily calculated by choosing
$\vc\theta =(\theta,0)$, in which case $\gamma_{\rm t}=-\gamma_1$,
$\gamma_\times=-\gamma_2$, and expressing the shear in terms of its
Fourier modes,
\bea
\ave{\gamma(\vc 0)\gamma^*(\vc\theta)}&=&\xi_+(\theta)
\nonumber \\
=&&\!\!\!\!\!\!\!\!\!\int{\d^2\ell\over (2\pi)^2}
\int{\d^2\ell'\over (2\pi)^2}\,
{\rm e}^{{\rm i}\vc\ell'\cdot\vc\theta}
\ave{\hat\gamma(\vc\ell)\hat\gamma^*(\vc\ell')} 
\elabel{gg1}
\\
=&&\!\!\!\!\!\!\!\!\!\int_0^\infty
{\d\ell\,\ell\over 2\pi}\,{\rm J}_0(\ell\theta)
\eck{P_{\rm E}(\ell)+P_{\rm B}(\ell)}\;,
\nonumber \\
\ave{\gamma(\vc 0)\gamma(\vc\theta)}&=&
\xi_-(\theta)+2{\rm i}\xi_\times(\theta)
\nonumber \\
=&&\!\!\!\!\!\!\!\!\!\int{\d^2\ell\over (2\pi)^2}
\int{\d^2\ell'\over (2\pi)^2}\,
{\rm e}^{-{\rm i}\vc\ell'\cdot\vc\theta}
\ave{\hat\gamma(\vc\ell)\hat\gamma(\vc\ell')} 
\elabel{gg2}\\
=&&\!\!\!\!\!\!\!\!\!
\int_0^\infty{\d\ell\,\ell\over 2\pi}\,{\rm J}_4(\ell\theta)
\eck{P_{\rm E}(\ell)-P_{\rm B}(\ell)+2{\rm i}P_{\rm EB}(\ell)}\;.
\nonumber
\eea
Making use of the orthogonality of Bessel functions,
\be
\int_0^\infty \d\theta\;\theta\,{\rm J}_\nu(s\theta)\,{\rm
J}_\nu(t\theta) ={\delta_{\rm D}(s-t)\over t}\;,
\elabel{N14}
\ee
we can invert the relations (\ref{eq:gg1}) and (\ref{eq:gg2}) and
express the power spectra in terms of the correlation functions,
\bea
P_{\rm E}(\ell)&=&\pi\int_0^\infty\d\theta\,\theta\,
\eck{\xi_+(\theta){\rm J}_0(\ell\theta)
+\xi_-(\theta){\rm J}_4(\ell\theta)}\;,
\nonumber\\
P_{\rm B}(\ell)&=&\pi\int_0^\infty\d\theta\,\theta\,
\eck{\xi_+(\theta){\rm J}_0(\ell\theta)
-\xi_-(\theta){\rm J}_4(\ell\theta)}\;,
\elabel{N15}
\\
P_{\rm EB}(\ell)&=&2\pi\int_0^\infty\d\theta\,\theta\,
\xi_\times(\theta){\rm J}_4(\ell\theta)
\;,
\nonumber
\eea
Hence, we have now expressed the various power spectra in terms of the
directly observable correlation functions $\xi$. One notes that the
correlation functions $\xi_+$ and $\xi_-$ depend on both, the E- and
B-mode power spectra, whereas the cross-correlation $\xi_\times$
depends on the cross-power $P_{\rm EB}$ only. It is obvious that the
cross-power and its corresponding correlation function do not `mix in'
with the E- and B-mode; in addition, the cross-power vanishes if the
shear field is statistically invariant under parity transformations,
which leave $\gamma_{\rm t}$ unchanged, but transform
$\gamma_\times \to -\gamma_\times$. One can therefore assume that in
realistic cases, $\xi_\times\equiv 0\equiv P_{\rm EB}$. However, since
cosmic shear is measured from finite data fields, cosmic variance may
lead to a non-zero measurement of the cross-power; in fact, the
measurement of the cross-power may serve as a lower limit on error bars
of the other power spectra.  
 
For a determination of the power spectra, the expressions
(\ref{eq:N15}) require a measurement of the correlation functions over
an infinite range in angle; whereas the
correlation functions decrease with $\theta$ and become very small for
large $\theta$, so that in effect the integrals can be replaced by
ones over a finite range of integration, one might want to obtain more
local decompositions into E- and B-modes.

\subsection{E/B-mode correlation functions}
We define the four correlation functions
\bea
\xi_{\rm E,B+}(\theta)&=&\int{\d\ell\,\ell\over 2\pi}\;P_{\rm E,B}(\ell)\,{\rm
J}_0(\theta\ell) \;,
\nonumber \\
\xi_{\rm E,B-}(\theta)&=&\int{\d\ell\,\ell\over 2\pi}\;P_{\rm E,B}(\ell)\,{\rm
J}_4(\theta\ell) \;,
\elabel{corrfcn}
\eea
which are defined such that in the absence of B-modes, $\xi_{\rm
E\pm}\equiv \xi_\pm$; these four correlation functions have also been
defined in C01, although only the `+' ones were investigated in more
detail there. Inserting (\ref{eq:N15}) into the
foregoing definitions, one obtains
\bea
\xi_{\rm E+}(\theta)&=&
{1\over 2}\int_0^\infty\d\ell\,\ell\int_0^\infty\d\vt\,\vt
{\rm J}_0(\theta\ell)
\nonumber \\
&\times& \eck{\xi_+(\vt){\rm J}_0(\vt\ell)
+\xi_-(\vt){\rm J}_4(\vt\ell)}\;.
\elabel{N17}
\eea
The $\ell$-integration can be carried out; consider the function
\be
G(\vt,\theta)=\int_0^\infty\d t\; t\,{\rm J}_0(t\vt)\,{\rm
J}_4(t\theta)\;.
\elabel{N18}
\ee
Making use of the recurrence relations for Bessel functions, one
can express ${\rm J}_4$ as
\[
{\rm J}_4(x)={24\over x^2}{\rm J}_2(x)-{8\over x}{\rm J}_1(x)
+{\rm J}_0(x)\;.
\]
By using Eq.\ts (11.4.41) of Abramowitz \& Stegun (1965), together with
(\ref{eq:N14}), one can perform the integration in (\ref{eq:N18}) term by
term to obtain
\be
G(\vt,\theta)=\rund{{4\over\theta^2}-{12\vt^2\over \theta^4}}{\rm
H}(\theta-\vt) +{1\over\theta}\delta_{\rm D}(\theta-\vt)\;,
\elabel{N19}
\ee
where ${\rm H}(x)$ is the Heaviside step function. We also note the
interesting property,
\bea
&&\int_0^\infty\d\vt\;\vt\,G(\vt,\theta)\,G(\vt,\vp)
={\delta_{\rm D}(\theta-\vp)\over \vp} 
\nonumber \\
&&=\int_0^\infty\d\vt\;\vt\,G(\theta,\vt)\,G(\vp,\vt)\, ,
\eea
which is readily shown using (\ref{eq:N14}).
Thus,
(\ref{eq:N17}) becomes
\bea
&&\xi_{\rm E+}(\theta)={1\over 2}\eck{\xi_+(\theta)
+\int_0^\infty\d\vt\,\vt\,\xi_-(\vt)\,G(\theta,\vt)}
\nonumber \\
&&={1\over 2}\eck{\xi_+(\theta)+\xi_-(\theta)
+\int_\theta^\infty{\d\vt\over\vt}\xi_-(\vt)\rund{4-12{\theta^2\over \vt^2}}}.
\elabel{xi1}
\eea
We have obtained a combination of shear correlation functions
which depends only on the E-modes; however, in order to obtain
$\xi_{\rm E+}$ one would need to know $\xi_-$ for arbitrarily large 
separations. However, 
\bea
&&\xi_{\rm E-}(\theta)={1\over 2}\eck{\xi_-(\theta)
+\int_0^\infty\d\vt\,\vt\,\xi_+(\vt)\,G(\vt,\theta)}
\nonumber \\
&&={1\over 2}\eck{\xi_-(\theta)+\xi_+(\theta)
+\int_0^\theta{\d\vt\,\vt\over\theta^2}\xi_+(\vt)
\rund{4-12{\vt^2\over \theta^2}}}\,.
\elabel{xi2}
\eea
depends on the observable correlation functions $\xi_\pm$ over a
finite range only and thus can be measured from finite data sets. 
Analogously, we find for the B-mode correlation functions
\bea
&&\xi_{\rm B+}(\theta)={1\over 2}\eck{\xi_+(\theta)
-\int_0^\infty\d\vt\,\vt\,\xi_-(\vt)\,G(\theta,\vt)}
\nonumber \\
&&={1\over 2}\eck{\xi_+(\theta)-\xi_-(\theta)
-\int_\theta^\infty{\d\vt\over\vt}\xi_-(\vt)
\rund{4-12{\theta^2\over \vt^2}}}\;,
\nonumber \\
&&\xi_{\rm B-}(\theta)={1\over 2}\eck{-\xi_-(\theta)
+\int_0^\infty\d\vt\,\vt\,\xi_+(\vt)\,G(\vt,\theta)}
\elabel{xi3}
\\
&&={1\over 2}\eck{\xi_+(\theta)-\xi_-(\theta)
+\int_0^\theta{\d\vt\,\vt\over\theta^2}\xi_+(\vt)
\rund{4-12{\vt^2\over \theta^2}}} \,. \nonumber
\eea
We note that $\xi_{\rm E+}+\xi_{\rm B+}=\xi_+$, $\xi_{\rm E-}-\xi_{\rm
B-}=\xi_-$. In order to calculate the E- and B-mode correlation
functions, one needs to know either the observable correlation
function $\xi_-$ to arbitrarily large, or $\xi_+$ to arbitrarily small
separations. This is of course impossible, owing to the finite size of
data fields on the one hand, and the impossibility to measure shapes
of very close pairs of galaxies. In either case, the lack of
measurements for large (or small) separations can be summarized in two
constants: suppose that $\xi_+$ can be measured down to separations of
$\theta_{\rm min}$; then, the integral in the `$-$' modes over $\xi_+$
can be split into one from 0 to $\theta_{\rm min}$, and one from
$\theta_{\rm min}$ to $\theta$. The former one has the
$\theta$-dependence $a/\theta^2-b/\theta^4$, where $a$, $b$ are two
constants, depending on $\xi_+$ for $\theta < \theta_{\rm min}$. The
decline of this contribution, with leading order $\theta^{-2}$, shows
that it has a small influence on the determination of the `$-$' modes
for $\theta\gg \theta_{\rm min}$; in addition, `reasonable guesses'
for $a$ and $b$ may be obtained by extrapolating the measured
$\xi_+$ towards small angles. The same reasoning shows the analogous
situation for the `+' modes.

\subsection{Aperture measures}
One very convenient way to separate E- and B-modes is provided by the
aperture mass: Defining the tangential and cross component of the shear
relative to the center of a circular aperture of angular radius
$\theta$, and defining
\bea
M_{\rm ap}(\theta)&=&\int\d^2\vt\;Q(|\vc\vt|)\,\gamma_{\rm t}(\vc\vt)\;,
\nonumber \\
M_{\perp}(\theta)&=&\int\d^2\vt\;Q(|\vc\vt|)\,\gamma_{\times}(\vc\vt)\;,
\elabel{N24}
\eea
it was shown in C01 that E-modes do not contribute to $M_\perp$, and
B-modes do not contribute to $M_{\rm ap}$; in fact, this can be easily
seen directly by inserting the Fourier transform of the shear
(\ref{eq:N10}) into (\ref{eq:N24}). Here, $Q(\vt)$ is an
axially-symmetric weight function which can be chosen arbitrarily. The
integration range in the foregoing equations extends over the support
of the weight function $Q$.  The aperture mass was introduced by
Schneider (1996) in an attempt to detect mass peaks from shear fields,
and later used by SvWJK as a two- and three-point statistics for
cosmic shear. $M_{\rm ap}$ can also be written as a filtered version
of the surface mass density (with a different and compensated weight
function which is related to $Q$), whereas $M_\perp$ has no direct
physical interpretation; in the absence of B-modes, $M_\perp$ should
vanish, and any non-vanishing signal is usually interpreted as being
due to noise or remaining systematics, and thus as a convenient error
estimate for $M_{\rm ap}$.  In SvWJK, a family of convenient weight
functions $Q$ was considered, the simplest of which is
\[
Q(\vt)={6\over \pi\theta^2}\,{\vt^2\over\theta^2}
\rund{1-{\vt^2\over\theta^2}}\,{\rm H}(\theta-\vt)\;,
\]
where $\theta$ is the radius of the aperture. This form of the weight
function shall be assumed in the following.

Using the complex number 
\bea
M(\theta)&=&M_{\rm ap}(\theta)+{\rm i}M_\perp(\theta)
\nonumber \\
&=&-\int\d^2\vt\;Q(|\vc\vt|)\,\gamma(\vc\vt)\,{\rm e}^{-2{\rm i}\vp}\;,
\nonumber
\eea
where $\vp$ is the polar angle of $\vc\vt$, one finds that 
\bea
\ave{M_{\rm ap}^2}\!&+&\!\ave{M_\perp^2}=\ave{MM^*}
\nonumber \\
&=&{1\over 2\pi}\int_0^\infty\d\ell\;\ell\,\eck{P_{\rm E}(\ell)+P_{\rm
B}(\ell)} W(\theta\ell)\;,
\elabel{N26}
\eea
with
\be
W(\eta):={576{\rm J}_4^2(\eta)\over \eta^4}\;,
\elabel{N26a}
\ee
which was derived by using the Fourier transform (\ref{eq:N10}) of the
shear, and the final steps are as in SvWJK. Similarly, one obtains
\bea
&&\ave{M_{\rm ap}^2}-\ave{M_\perp^2}+2{\rm i}\ave{M_{\rm ap}M_\perp}
=\ave{MM}
\nonumber \\
&&={1\over 2\pi}\int_0^\infty\d\ell\;\ell\,\eck{P_{\rm E}(\ell)-P_{\rm
B}(\ell)+2{\rm i}P_{\rm EB}(\ell)} 
W(\theta\ell)\;.
\elabel{N27}
\eea
Combining the two previous equations, one thus gets
\bea
\ave{M_{\rm ap}^2}(\theta)&=&
{1\over 2\pi}\int_0^\infty\d\ell\;\ell\,P_{\rm E}(\ell) \, 
W(\theta\ell)\;,
\nonumber \\
\ave{M_{\perp}^2}(\theta)&=&
{1\over 2\pi}\int_0^\infty\d\ell\;\ell\,P_{\rm B}(\ell) \, 
W(\theta\ell)\;,
\elabel{N28}
\eea
so that these two-point statistics clearly separate E- and B-modes. In
addition, they provide a highly localized measure of the corresponding
power spectra, since the filter function $W(\eta)$ 
involved is quite narrow (see SvWJK); 
in fact, Bartelmann \& Schneider (1999) have shown that replacing $W(\eta)$
in (\ref{eq:N28}) by $A \delta_{\rm D}(\eta-\eta_0)$, with $\eta_0\approx
4.25$, provides a fairly accurate approximation.
Furthermore, (\ref{eq:N27}) can
be used to check whether the shear data contain a contribution from
the cross power $P_{\rm EB}$.

The aperture measures can be obtained directly from the observational
data by laying down a grid of points, at each of which $M_{\rm ap}$
and $M_\perp$ are calculated from (\ref{eq:N24}). However, obtaining
the dispersion with this strategy turns out to be difficult in
practice, since data fields usually contain holes and gaps,
e.g. because of masking (for bright stars), bad columns etc. It is
therefore interesting to calculate these dispersions directly in terms
of the correlation functions, which can be done by inserting
(\ref{eq:N15}) into (\ref{eq:N28}),
\bea
\ave{M_{\rm ap}^2}(\theta)\!\! 
&=&\!\!{1\over
2}\int{\d\vt\,\vt\over\theta^2}
\eck{\xi_+(\vt)\,T_+\!\!\rund{\vt\over\theta}+\xi_-(\vt)\,T_-\!\!\rund{\vt\over\theta}}
\;,\nonumber \\
\ave{M_{\perp}^2}(\theta)\!\!&=&\!\!{1\over
2}\int{\d\vt\,\vt\over\theta^2}
\eck{\xi_+(\vt)\,T_+\!\!\rund{\vt\over\theta}-\xi_-(\vt)\,T_-\!\!\rund{\vt\over\theta}}
\;,\nonumber\\
\elabel{N29}
\eea
where we have defined the functions
\bea
T_+(x)&=&576\int_0^\infty{\d t\over t^3}\,{\rm J}_0(x t)\,\eck{{\rm
J}_4(t)}^2 \;,\nonumber \\
T_-(x)&=&576\int_0^\infty{\d t\over t^3}\,{\rm J}_4(x t)\,\eck{{\rm
J}_4(t)}^2 \;.
\elabel{N30}
\eea
The integration range in (\ref{eq:N29}) formally extends from zero to
infinity, but as we shall see shortly, the functions $T_\pm(x)$ vanish for
$x>2$, so the integration range is $0\le \vt\le 2\theta$:
For $T_-(x)$, an analytic expression can be obtained, using Eq.\
(6.578.9) of Gradshteyn \& Ryzhik (1980),
\be
T_-(x)={192\over 35\pi}x^3\rund{1-{x^2\over 4}}^{7/2}\,{\rm H}(2-x)\;,
\elabel{N31}
\ee
so that $T_-(x)$ vanishes for $x>2$.
Furthermore, the two functions $T_+$ and $T_-(x)$ are related: using
(\ref{eq:N14}), one finds that
\[
\int_0^\infty\d x\;x\,T_+(x)\,{\rm J}_0(t x)
=W(t)=
\int_0^\infty\d x\;x\,T_-(x)\,{\rm J}_4(t x)
\]
so that 
\bea
T_-(x)&=&\int_0^\infty\d t\;t\,W(t)\,{\rm J}_4(x t)
=\int_0^\infty \d y\;y\,T_+(y)\,G(y,x)\; ,
\nonumber \\
T_+(x)&=&\int_0^\infty\d t\;t\,W(t)\,{\rm J}_0(x t)
=\int_0^\infty \d y\;y\,T_-(y)\,G(x,y)\; ;
\nonumber 
\eea
using the latter expression, together with (\ref{eq:N31})
and (\ref{eq:N19}), one obtains  
\bea
&&T_+(x)={6(2-15x^2)\over 5}\eck{1-{2\over\pi}\arcsin\rund{x\over 2}}
\nonumber \\
&&+ {x\sqrt{4-x^2}\over 100\pi}
\rund{120+2320x^2-754x^4+132 x^6-9x^8}
\elabel{N33}
\eea
for $x\le 2$, and $T_+(x)$ vanishes for $x>2$. Hence, the integrals in
(\ref{eq:N29}) extend only over $0\le \vt\le 2\theta$, so that
$\ave{M_{\rm ap}^2}$ and $\ave{M_\perp^2}$ can be obtained directly in
terms of the observable correlation function $\xi_\pm$ over a finite
interval. The two functions $T_\pm$ are plotted in Fig.\ts 1.

   \begin{figure}
   \includegraphics[width=9cm]{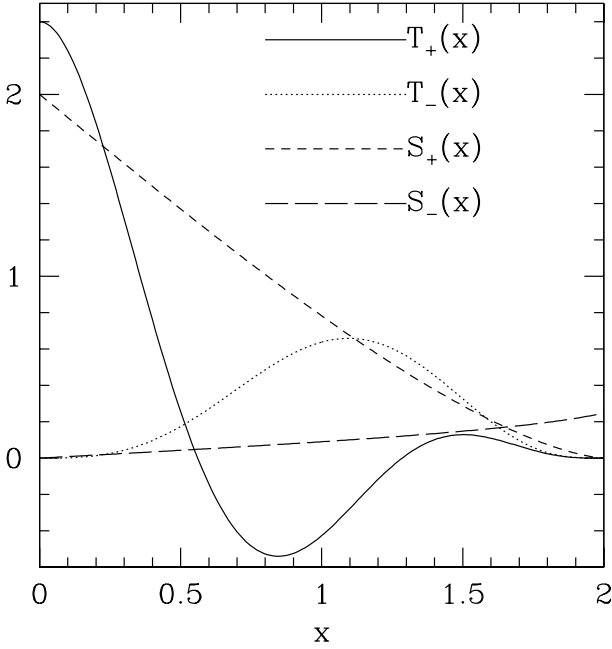}
   \caption{The four functions defined in text}
    \end{figure}

\subsection{Shear dispersion}
Another cosmic shear statistics often employed is the shear dispersion
in a circle of angular radius $\theta$. It is related to the power
spectra by
\be
\ave{\abs{\bar\gamma}^2}(\theta)={1\over 2\pi}\int\d\ell\,\ell\,
(P_{\rm E}+P_{\rm B})(\ell)\,W_{\rm TH}(\ell\theta)\;,
\elabel{Q1}
\ee
where
\be
W_{\rm TH}(\eta)={4 {\rm J}_1^2(\eta)\over \eta^2}
\ee
is the top-hat filter function. In contrast to the aperture measures
of the previous subsection, the shear dispersion (\ref{eq:Q1})
contains both modes; furthermore, the filter function $W_{\rm
TH}(\eta)$ is much broader than $W(\eta)$ in (\ref{eq:N26a}), as
demonstrated in SvWJK. It thus provides a much less localized measure
of the power spectra than the aperture measures. On the other hand,
this larger filter width implies that the signal of the shear
dispersion is larger than that of the aperture measures, which
explains why the first cosmic shear detections (van Waerbeke et al.\
2000; Bacon et al.\ 2001, Kaiser et al.\ 2000) were obtained in terms
of the shear dispersion.

As before, the shear dispersion can be obtained by calculating the
mean shear in circles which are laid down on a grid of points, with
the drawback of being affected by gaps in the data
field. Alternatively, the shear dispersion can be obtained directly
from the correlation function,
\be
\ave{\abs{\bar\gamma}^2}(\theta)=\int{\d\vt\,\vt\over\theta^2}\,
\xi_+(\vt)\,S_+\rund{\vt\over\theta}\;,
\elabel{Q2}
\ee
where (van Waerbeke 2000)
\be
S_+(x)={1\over\pi}\eck{4\arccos\rund{x\over 2}-x\sqrt{4-x^2}}
\ee
for $x\le 2$, and zero otherwise. Hence, the integral in (\ref{eq:Q2})
extends only over the finite interval $0\le\vt\le 2\theta$, which
makes this a convenient way to calculate the shear dispersion. 

One can also define the shear dispersions of the E- and B-mode,
according to 
\be
\ave{\abs{\bar\gamma}^2}_{\rm E,B}(\theta)={1\over 2\pi}\int\d\ell\,\ell\,
P_{\rm E,B}(\ell)\,W_{\rm TH}(\ell\theta)\;,
\elabel{Q3}
\ee
but they cannot be individually obtained from measuring the shear
directly. Nevertheless, both of these dispersions can be obtained in
terms of the correlation functions,
\bea
\ave{\abs{\bar\gamma}^2}_{\rm E,B}\!\!(\theta)\!\!&=&\!\!
\int\!\!{\d\vt\,\vt\over 2 \theta^2}
\eck{\xi_+(\vt)\,S_+\!\!\rund{\vt\over\theta}
\pm\xi_-(\vt)\,S_-\!\!\rund{\vt\over\theta}} \;,
\nonumber\\
\elabel{Q4}
\eea
which can be derived in close analogy to the derivation of
(\ref{eq:N29}), and the function $S_-$ is related to $S_+$ in the same
way as the corresponding $T$-functions,
\bea
S_-(x)&=&\int_0^\infty \d y\;y\,S_+(y)\,G(y,x)\nonumber \\
&=&{x\sqrt{4-x^2}(6-x^2)-8(3-x^2)\arcsin(x/2)
\over \pi x^4}
\eea
for $x\le 2$, and $S_-(x)=4(x^2-3)/x^4$ for $x>2$. Hence, the
integrals in (\ref{eq:Q4}) do not cut off at finite separation, which
was to be expected, since a constant shear cannot be uniquely assigned
to an E- or B-mode, but contributes to $ \ave{\abs{\bar\gamma}^2}$.

\section{B-mode from source clustering}
In the previous section we have presented the decomposition of a
general shear field into E/B-modes. It is usually assumed that lensing
alone yields a pure E-mode shear field, so that the detection of a
B-mode in the van Waerbeke et al.\ (2001) data (see also Pen et al.\
2002) was surprising and interpreted as being due to systematic errors
or a signature of intrinsic alignment of sources. Here we show that
lensing indeed {\it does} generate a B-mode component of the shear if
the source galaxies from which the shear is measured are clustered.

\subsection{Correlation functions and power spectra}
Define the `equivalent' surface mass density for a fixed source
redshift, or comoving distance $w$,
\be
\kappa(\vc\theta,w)=
\int_0^w \d w'\; F(w',w)\;
\delta[f(w')\vc\theta,w']\;.
\elabel{1}
\ee
where
\be
F(w',w)={3 H_0^2 \Omega_0\over 2 c^2}\,{f(w') f(w-w')\over a(w')
f(w)}\;;
\ee
here, $H_0$ and $\Omega_0$ denote the Hubble constant and the density
parameter, $w$ is the comoving distance, $f(w)$ is the comoving
angular-diameter distance to comoving distance $w$, $\delta$ is the
density contrast, and $a(w)=(1+z)^{-1}$ is the cosmic scale factor,
defined such that $a=1$ today, again using the notation of BS01.
Accordingly, we define the shear components
\be
\gamma_\alpha(\vc\theta,w)= 
\rund{{\D_\alpha\over \pi} * \kappa}(\vc\theta,w)\;,
\elabel{2}
\ee
where the operator $\D$ is defined in (\ref{eq:1.1}) and
(\ref{eq:1.2}). 
Then, the shear correlation function for two sources at positions
$\vc\theta_i$ and distances $w_i$ becomes
\bea
&&\ave{\gamma_\alpha(\vc\theta_1,w_1)\,\gamma_\beta(\vc\theta_2,w_2)}
\!=\!\rund{\D_\alpha \D_\beta\over \pi^2} \!*\!
\int_0^{w_1} \!\!\!\d w_1'\;
F(w_1',w_1)
\nonumber \\ &&\times\!\!
\int_0^{w_2} \!\!\!\!\!\d w_2'\;
F(w_2',w_2)
\ave{\delta[f(w_1')\vc\theta_1,w_1']\;\delta[f(w_2')\vc\theta_2,w_2']},
\nonumber
\eea
where the first $\D$ operates on $\vc\theta_1$, and the second $\D$ on
$\vc\theta_2$. 
This equation is of the form (BS\ 2.78), and thus we obtain, using
(BS\ 2.82),
\bea
\ave{\gamma_\alpha\,\gamma_\beta}
&=&{\D_\alpha \D_\beta\over \pi^2} *
\int_0^{w_{1,2}} \d w\;
F(w,w_1)\,F(w,w_2)
\nonumber \\ &\times&\!\!\!
\int\!\!{\d^2 k\over (2\pi)^2}\,P_\delta(|\vc k|,w)
\exp\eck{{\rm i}f(w)\vc k\cdot (\vc\theta_1-\vc\theta_2)}\,,
\elabel{G55}
\eea
where we have temporarily dropped the arguments of the shear
correlator, $w_{1,2}=\min(w_1,w_2)$, and $P_\delta$ is the power
spectrum of the density fluctuations which develops as a function of
cosmic time (or as a function of comoving distance $w$). The upper
limit of the integral expresses the fact that a correlated shear can
only be generated by matter which is at smaller distance than both
sources.

The operators $\D$ only act on the final term of (\ref{eq:G55}) which
can be evaluated using the Fourier transform of $\D$, as in
(\ref{eq:N10}), 
\[
D_\alpha \D_\beta * {\rm e}^{{\rm i}f\vc
k\cdot(\vc\theta_1-\vc\theta_2)}
=
\hat\D_\alpha^*(f\vc k)\hat\D_\beta(f\vc k) {\rm e}^{{\rm i}f\vc
k\cdot(\vc\theta_1-\vc\theta_2)} 
\;,
\]
so that
\bea
&&\ave{\gamma_\alpha\,\gamma_\beta}
={9 H_0^4 \Omega_0^2\over 4 c^4}
\int_0^{w_{1,2}} {\d w\over a^2(w)}\;
{f(w_1-w)\over f(w_1)}
{f(w_2-w)\over f(w_2)}
\nonumber \\ &&\times
\int{\d^2 \ell\over (2\pi)^2}\,P_\delta\rund{{|\vc \ell|\over f(w)},w}
{ \hat\D_\alpha^*(\vc \ell)\hat\D_\beta(\vc \ell)\over \pi^2}\,
{\rm e}^{{\rm i}\vc
\ell\cdot(\vc\theta_1-\vc\theta_2)} \,.
\eea
Evaluating the relevant combinations, one finds
\bea
&&\rund{\ave{\gamma_{\rm t}\gamma_{\rm t}} +\ave{\gamma_\times
\gamma_\times}}(\theta;w_1,w_2)
=
{9 H_0^4 \Omega_0^2\over 4 c^4}
\int_0^{w_{1,2}} {\d w\over a^2(w)}\nonumber \\
&&\times
R(w,w_1)\,R(w,w_2)
 \int{\d \ell\,\ell\over (2\pi)}\,P_\delta\rund{{ \ell\over f(w)},w}\,{\rm
J}_0(\ell\theta) \,, \nonumber
\eea
\bea
&&\rund{\ave{\gamma_{\rm t}\gamma_{\rm t}} -\ave{\gamma_\times
\gamma_\times}}(\theta;w_1,w_2)
=
{9 H_0^4 \Omega_0^2\over 4 c^4}
\int_0^{w_{1,2}} {\d w\over a^2(w)}\nonumber \\
&&\times
R(w,w_1)\,R(w,w_2)
 \int{\d \ell\,\ell\over (2\pi)}\,P_\delta\rund{{ \ell\over f(w)},w}\,{\rm
J}_4(\ell\theta) \,,\nonumber
\elabel{8}
\eea
where $\theta=\abs{\vc\theta_1-\vc\theta_2}$ and
$R(w,w_i)=f(w_i-w)/f(w_i)$ is the ratio of the angular diameter
distances of a source at $w_i$ seen from the distance $w\le w_i$ and
that seen from an observer at $w=0$.

When measuring cosmic shear from source ellipticities, the source
galaxies have a broad distribution in redshift, unless information on
the redshifts are available and taken into account. Hence, to
calculate the observable shear correlation functions, the
foregoing expressions need to be averaged over the source redshift
distribution. Let $p(w_1,w_2;\theta)$ be the probability density for
comoving distances of two sources separated by an angle $\theta$ on
the sky; then we have for the observable correlation functions 
\be
\xi_\pm(\theta)=\int_0^{w_{\rm H}}\!\!\!\!\!
\d w_1\int_0^{w_{\rm H}}\!\!\!\!\!\d
w_2\;p(w_1,w_2;\theta)
\rund{ \ave{\gamma_{\rm t}\gamma_{\rm t}} \!\pm\! \ave{\gamma_\times
\gamma_\times}}
\elabel{9}
\ee
and $w_{\rm H}$ is the comoving distance to the horizon.
By changing the order of integration  according to
\[
\int_0^{w_{\rm H}}\!\!\! \d w_1\int_0^{w_{\rm H}}\!\!\!\d
w_2\int_0^{w_{1,2}}\!\!\!\d w
=\int_0^{w_{\rm H}}\!\!\!\d w\int_w^{w_{\rm H}}\!\!\!
\d w_1\int_w^{w_{\rm H}}\!\!\!\d w_2
\]
we obtain
\bea
\xi_\pm(\theta)
&=&
{9 H_0^4 \Omega_0^2\over 4 c^4}
\int_0^{w_{\rm H}} {\d w\over a^2(w)}\;
\int{\d \ell\,\ell \over (2\pi)}\,P_\delta\rund{{ \ell\over f(w)},w}
\nonumber \\
&\times&
{\rm
J}_{0,4}(\ell\theta)
\ave{R(w,w_1)\,R(w,w_2)} \;,
\elabel{G56}
\eea
where the angular brackets denote the averaging of the
angular-diameter distance ratios over the source distance
distribution. 
We shall write the source redshift distribution as
\be
p(w_1,w_2;\theta)={\bar p(w_1)\,\bar p(w_2)\eck{1+\delta_{\rm
D}(w_1-w_2)
g(w_1;\theta)} \over 1+\omega(\theta)} \,,
\elabel{12}
\ee
where
\be
\omega(\theta)=\int_0^{w_{\rm H}}\d w\;\bar p^2(w)\,g(w;\theta)
\elabel{angcor}
\ee
is the angular correlation function of the galaxies, and $\bar p(w)$
describes their redshift distribution. The second term in
(\ref{eq:12}) accounts for source clustering.  In making this ansatz,
we have accounted for the fact that redshift clustering occurs only
over a very small interval in redshift over which all the other
redshift-dependent functions occurring in (\ref{eq:G56}) can be
considered constant. Note that (\ref{eq:12}) is normalized,
\[
\int\d w_1\int\d w_2 \; p(w_1,w_2;\theta) =1\;,
\]
as required. Then, the average of the angular-diameter distance
ratios becomes
\be
\ave{R(w,w_1)\,R(w,w_2)}={\bar W^2(w)+V(w,\theta)\over 1+\omega(\theta)}\;,
\elabel{RR}
\ee
where
\be
\bar W(w)=\int_w^{w_{\rm H}}\d
w_1\;\bar p(w_1) {f(w_1-w)\over f(w_1)}\;,
\ee
\be
V(w,\theta)=\int_w^{w_{\rm H}}\d w_1\;\bar p^2(w_1)
\rund{f(w_1-w)\over f(w_1)}^2\,g(w_1;\theta)\;.
\ee
The correlation of sources thus yields an average of the
angular-diameter distance ratios which is not simply the square of the
mean distance ratio $\bar W$, but contains in addition a correlated
part described by $V$ and the normalization correction $1+\omega$. If
the angular separation of the sources is large, the correlation in
redshift is expected to be small; hence, for large separations one
expects $V$ and $\omega$ to vanish. The degree of redshift correlation
depends on the angular separation considered; the fact that the mean
of the product of the angular diameter distance ratio (\ref{eq:RR})
depends on the separation $\theta$ is the cause for a B-mode
contribution to the shear correlation function!

One can check that the correlated redshift probability distribution
behaves as expected in some simple cases. For example, if $w\ll c/H_0$
is very much smaller than the characteristic source distance $w_0$,
one finds that
\[
\ave{R(w,w_1)\,R(w,w_2)}\approx 1\;;
\]
in this case, the lensing strength of matter at distance $w$ is
basically independent of the exact source redshift, so that source
redshift clustering is irrelevant for those lens redshifts. Another
case of interest occurs when the selected sources come from a very
narrow distance interval, of width $\Delta w$ centered on $w_0$; then,
(\ref{eq:angcor}) yields the relation $\omega(\theta)\approx
g(w_0;\theta)/\Delta w$, and 
\[
\ave{R(w,w_1)\,R(w,w_2)}\approx\eck{f(w_0-w)\over f(w_0)}^2\,{\rm
H}(w_0-w)\;.
\]
Hence, also in this case, $\ave{RR}$ does not depend on $\theta$, and
therefore no B-mode contribution occurs -- as noted before, if all
sources are at the same redshift, one obtains a pure E-mode shear
field.

We can now rewrite (\ref{eq:G56}) in the form
\be
\xi_\pm(\theta)={1\over 1+\omega(\theta)}
\int{\d \ell\;\ell\over (2\pi)}\,
{\rm J}_{0,4}(\ell\theta) \eck{P_\kappa(\ell)+P_{\rm
c}(\ell;\theta)}
\;,
\elabel{G60}
\ee
where the 0 (4) corresponds to $\xi_+$ ($\xi_-$), 
\be
P_\kappa(\ell)=
{9 H_0^4 \Omega_0^2\over 4 c^4}
\int_0^{w_{\rm H}} {\d w\over a^2(w)}\,\bar
W^2(w)\,P_\delta\rund{{\ell\over f(w)},w}\;,
\elabel{G61}
\ee
and
\be
P_{\rm c}(\ell;\theta)={9 H_0^4 \Omega_0^2\over 4 c^4}
\int_0^{w_{\rm H}} {\d w\over a^2(w)} V(w,\theta)
\,P_\delta\rund{{\ell\over
f(w)},w} \;.
\elabel{G62}
\ee
The first term of (\ref{eq:G60}) in the absence of source correlations
(i.e., $\omega=0$) is the one usually derived in cosmic shear
considerations; $P_\kappa$ is the power spectrum of the projected
matter density, related to the three-dimensional power spectrum
$P_\delta$ by a Limber-type equation (e.g., Kaiser 1992). The second
term in (\ref{eq:G60}) and the `normalization correction' $1+\omega$
comes about due to source correlations.

\subsection{The E/B-mode decomposition}
From the correlation functions (\ref{eq:G60}), by writing
\bea
{P_\kappa(\ell)+P_{\rm c}(\ell;\theta)\over 1+\omega(\theta)}
&=&P_\kappa(\ell)+\P(\ell;\theta)\nonumber \\
&\equiv& P_\kappa(\ell) +{P_{\rm
c}(\ell;\theta)-\omega(\theta)P_\kappa(\ell) \over
1+\omega(\theta)}\;,
\eea
we can derive the
E- and B-mode power spectra, making use of (\ref{eq:N15}),
\be
P_{\rm E}(\ell)=P_\kappa(\ell)+ P_{\rm cE}(\ell)\; ;
\;\; P_{\rm B}(\ell)=P_{\rm cB}(\ell)\; ;
\ee
with
\bea
P_{\rm cE,B}(\ell)&=&
{1\over 2}\int\d\theta\;\theta
\int\d \ell' \, \ell'\,\P(\ell';\theta)
\nonumber\\
&\times&
\eck{{\rm J}_0(\ell\theta){\rm J}_0(\ell'\theta)
   \pm{\rm J}_4(\ell\theta){\rm J}_4(\ell'\theta)}\;,
\elabel{G65}
\eea
where the `+' (`$-$')-sign corresponds to the E- (B-) mode.  If the
ratio containing the power spectra did not depend on $\theta$,
$P_{\rm B}$ would vanish identically, as it should. However, this term
{\it does} depend on $\theta$ due to source correlations; therefore,
without using redshift information, the presence of B-modes in cosmic
shear observations is unavoidable.

Using the definitions of the E- and B-mode correlation function, we
obtain 
\bea
\xi_{\rm E+}(\theta)&=&\int{\d \ell\;\ell\over (2\pi)}\,P_\kappa(\ell)\,
{\rm J}_0(\ell\theta)\nonumber \\
&+&\int{\d \ell\;\ell\over (4\pi)}\Biggl[\P(\ell;\theta)
\eck{{\rm J}_0(\ell\theta)+{\rm J}_4(\ell\theta)} \\
&+&\int_\theta^\infty\d\vt\,\vt\,\P(\ell;\vt){\rm J}_4(\ell\vt)
\rund{{4\over \vt^2}-{12\theta^2\over \vt^4}}\Biggr]\;, \nonumber
\eea
\bea
\xi_{\rm B+}(\theta)&=&\int{\d \ell\;\ell\over (4\pi)}
\Biggl[\P(\ell;\theta)
\eck{{\rm J}_0(\ell\theta)-{\rm J}_4(\ell\theta)} \nonumber \\
&-&\int_\theta^\infty\d\vt\,\vt\,\P(\ell;\vt){\rm J}_4(\ell\vt)
\rund{{4\over \vt^2}-{12\theta^2\over \vt^4}}\Biggr]\;,
\eea
\bea
\xi_{\rm E-}(\theta)&=&\int{\d \ell\;\ell\over (2\pi)}\,P_\kappa(\ell)\,
{\rm J}_4(\ell\theta)\nonumber \\
&+&\int{\d \ell\;\ell\over (4\pi)}\Biggl[\P(\ell;\theta)
\eck{{\rm J}_0(\ell\theta)+{\rm J}_4(\ell\theta)} \\
&+&\int_0^\theta \d\vt\,\vt\,\P(\ell;\vt){\rm J}_0(\ell\vt)
\rund{{4\over \theta^2}-{12\vt^2\over \theta^4}}\Biggr]\;, \nonumber
\eea
\bea
\xi_{\rm B-}(\theta)&=&\int{\d \ell\;\ell\over (4\pi)}
\Biggl[\P(\ell;\theta)
\eck{{\rm J}_0(\ell\theta)-{\rm J}_4(\ell\theta)} \nonumber \\
&+&\int_0^\theta\d\vt\,\vt\,\P(\ell;\vt){\rm J}_0(\ell\vt)
\rund{{4\over \theta^2}-{12\vt^2\over \theta^4}}\Biggr]\;.
\eea

\section{Relative strength of the B-mode}
Whereas the presence of a B-mode, and an additional contribution to
the E-mode due to source clustering must occur, one needs to estimate
the relative amplitude of this effect as compared to the `usual'
cosmic shear strength described by $P_\kappa$. This estimate requires
a model for the source clustering, i.e., a model for the function
$g(w;\theta)$. $g$ can be related to the three-dimensional correlation
function $\xi_{\rm gg}$ of galaxies,
\be
g(w;\theta)=\int_{-\infty}^\infty \d(\Delta w)\;
\xi_{\rm gg}\rund{\sqrt{ (\Delta w)^2+f^2(w)\theta^2}}\;.
\ee
If we assume that the correlation function behaves like a power-law,
$\xi_{\rm gg}(r)=[r/r_0(w)]^{-\gamma}$, where $r_0(w)$ is the comoving
correlation length, and $\gamma\simeq 1.7$, then
\be
g(w;\theta)=C\,r_0(w)\rund{f(w)\theta\over r_0(w)}^{1-\gamma}\;,
\ee
where
$C={\sqrt{\pi}\,\Gamma([\gamma-1]/2)/\Gamma(\gamma/2)}$,
and $\Gamma(x)$ is the Gamma function. This yields for the angular
two-point correlation function
\be
\omega(\theta)=\theta^{(1-\gamma)} C\int_0^{w_{\rm H}}\d w\;\bar
p^2(w)\,r_0(w)\,\rund{f(w)\over r_0(w)}^{1-\gamma} \;.
\ee
A useful parameterization of $\omega(\theta)$ is 
$\omega=A(1') (\theta/1')^{(1-\gamma)}$. Fixing $\gamma=1.7$, one
obtains a relation between the correlation length $r_0(w)$ and the
redshift distribution $\bar p(w)$, 
\be
\int_0^{w_{\rm H}}\d w\;\bar
p^2(w)\,r_0(w)\,\rund{f(w)\over r_0(w)}^{-0.7}
=1.65\times 10^{-5}\,{A(1')\over 0.02} \;,
\elabel{64}
\ee
where the fiducial value of $A(1')$ was taken from McCracken et al.\ts
(2001). Note that in McCracken et al.\ts (2001), essentially the same
data set has been used as in the cosmic shear analysis of van Waerbeke
et al.\ (2001); in particular, the depth of the data are the same. The
above-quoted value for the angular clustering strength at $1'$
corresponds to the faintest flux threshold considered in McCracken et
al., which is very similar to the flux limit employed in the cosmic
shear analysis. It must be mentioned, however, that the galaxies used
in the cosmic shear analysis do not form a truly flux-limited sample,
since additional cuts are used, e.g. a size cut. Hence, a precise
estimate of the angular correlation function of those galaxies which
were used for the cosmic shear analysis cannot be given. 

For this reason, we shall assume the power-law dependence of
$\omega(\theta)$ as given above; in addition, we will make the
simplifying assumption that the comoving clustering length $r_0(w)$ is
independent of distance $w$; this assumption is not too
critical, since the function $\bar p^2(w)$ is relatively
well peaked and therefore large $w$-variations of the correlation
length are not probed. Then, (\ref{eq:64}) determines this constant
comoving correlation length $r_0$. We obtain in this case
\[
g(w;\theta)=\omega(\theta){[f(w)]^{1-\gamma} \over
\int\d w'\,\bar p^2 (w')\,[f(w)]^{1-\gamma} }\;.
\]
The power-law dependence of $g$ on $\theta$ implies that $P_{\rm
c}(\ell;\theta)$ also behaves like $\theta^{1-\gamma}$. Since the
angular correlation function is small compared to unity, even on
scales of a few arcseconds, we shall neglect $\omega(\theta)$ in the
denominator of the integrand in (\ref{eq:G65}); this greatly
simplifies the calculation of the power spectra due to source
clustering, since the $\theta$-integration can then be carried out
first, making use of Eqs.\ts(11.4.33, 34) of AS.

In order to make further progress, we need to assume a redshift
distribution for the sources from which the shear is measured. We
employ the form (Brainerd et al. 1996)
\be
\bar p_z(z)\propto z^2 \exp\eck{-\rund{z\over z_0}^\beta}\;,
\ee
and shall consider $\beta=3/2$ in the following, yielding a mean
redshift of $\ave{z}\approx 1.5 z_0$.

   \begin{figure} 
   \includegraphics[width=9cm]{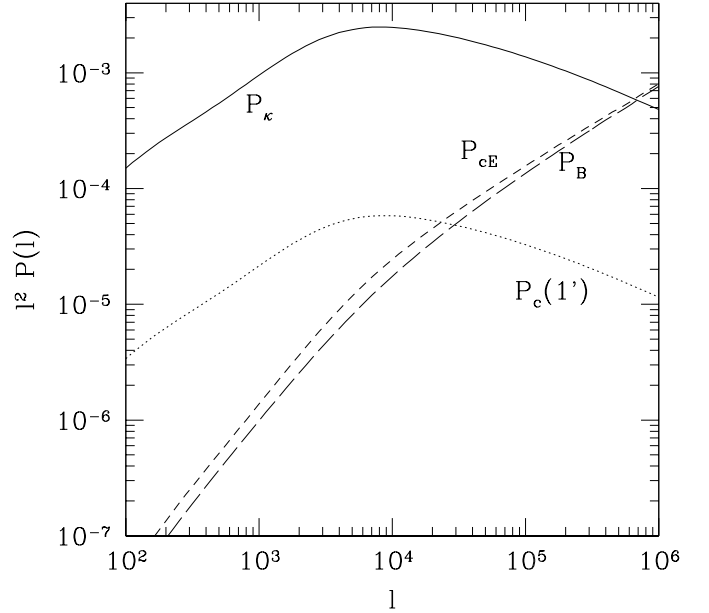}
   \caption{Dimensionless power spectra $\ell^2 P(\ell)$, as a
   function of of wavenumber $\ell$. The solid curve corresponds to
   the power spectrum $\ell^2 P_\kappa(\ell)$ that is the `standard'
   power spectrum of the projected mass density. The dotted curve
   displays $\ell^2 P_{\rm c}(\ell, 1')$, and the two dashed curves
   correspond to the E- and B-mode power caused by the source
   clustering. Here, a $\Lambda$CDM model was used, with shape
   parameter $\Gamma=0.21$, normalization $\sigma_8=1$, and the source
   redshift distribution is characterized by $z_0=1$, yielding
   $\ave{z}\approx 1.5$. Other parameters for the model used here are
   mentioned in the text.}  \end{figure}

In Fig.\ts 2 we show an example of the various power spectra
considered here; all power spectra are multiplied by $\ell^2$ to
obtain dimensionless quantities. For this figure, we employed a
standard $\Lambda$CDM model with $\Omega_\Lambda=0.7$ and
$\Omega_0=0.3$ and normalization $\sigma_8=1$. Sources are distributed
in redshift according to the foregoing prescription, with $z_0=1$. The
amplitude of the angular correlation function of galaxies was chosen
to be $A(1')=0.02$, and the slope of $\gamma=1.7$ for the
three-dimensional correlation function was used; the corresponding
correlation length in this case is $r_0\approx 4.7 h^{-1}{\rm Mpc}$.
To calculate the three-dimensional power spectrum and its redshift
evolution, we used the Peacock \& Dodds (1996) prescription for the
non-linear evolution of $P_\delta(k,w)$.

The power spectrum of the projected mass density, $P_\kappa(\ell)$, is
the same as that in SvWJK, except for a slightly different choice of
the cosmological parameters. The spectrum $P_c(\ell;1')$ is very much
smaller than $P_\kappa(\ell)$, as expected from the smallness of the
amplitude of the angular correlation function; in fact, the ratio 
$ P_c(\ell;1')/ P_\kappa(\ell)$ is nearly constant at a value of
approximately $(1+B)A(1')$, with $B\approx 1.2$ for this choice of the
parameters. 

The behavior of the power spectra which arise from source clustering,
$P_{\rm cE}$ and $P_{\rm B}\equiv P_{\rm cB}$, as a function of $\ell$
is quite different. First, both of these spectra are very similar,
which is due to the fact that the ${\rm J}_4$-term in (\ref{eq:G65})
is much smaller than the ${\rm J}_0$-term. Second, although both of
these spectra are small on large angular scales, i.e. at small $\ell$,
their relative value increases strongly for smaller angular
scales. Hence, as expected, the relative importance of source
clustering increases for larger $\ell$. What is surprising, though, is
that these power spectra have the same amplitude as $P_\kappa$ at a
value of $\ell\sim 6.7\times 10^5$, corresponding to an angular scale
of $\theta = 2\pi/\ell\sim 2''$, and the relative contribution of the
B-mode amounts to about 2\% at an angular scale of $1'$. It should be
noted here that cosmic shear has already been measured on scales below
$1'$; therefore, source clustering gives rise to a B-mode component in
cosmic shear which is observable.

We shall now consider the behavior of $P_{\rm B}(\ell)$ for large
values of $\ell$. The aforementioned properties of $P_{\rm
c}(\ell;\theta)$ can be summarized as
\[
P_{\rm c}(\ell;\theta)\approx (1+B) A(1')\,P_\kappa(\ell)
\rund{\theta\over 1'}^{1-\gamma}\;.
\]
Inserting this result into (\ref{eq:G65}), neglecting the ${\rm
J}_4$-terms and considering the limit $\ell\to \infty$ yields
\[
\ell^2P_{\rm B}(\ell)\approx a_1\, B\, A(1')\, (\ell 1')^{\gamma-1}
\int\d\ell\,\ell\,P_\kappa(\ell)\;,
\]
where $a_1=2^{1-\gamma}\Gamma[(3-\gamma)/2]/\Gamma[(\gamma-1)/2]
\approx 0.335$ for $\gamma=1.7$. Hence, $P_{\rm B}\propto 
\ell^{\gamma-3}$ at large $\ell$. To obtain an approximate value for
the integral in the preceding equation, we shall describe the power
spectrum $P_\kappa$ by a simple function,
\[
\ell^2 P_\kappa(\ell)\sim B_{\rm m}{ (\ell/\ell_{\rm m})^\alpha
\over \eck{1+(\ell/\ell_{\rm m})^2}^\beta}
\]
with $B_{\rm m}\sim 3.7\times 10^{-3}$, $\alpha\sim 0.7$, $\beta\sim
0.6$, and $\ell_{\rm m}\sim 7\times 10^3$. Using $A(1')=0.02$, this then
yields
\[
\ell^2 P_{\rm B}(\ell)\approx 1.5\times 10^{-4}
\rund{\ell\over 10^5}^{0.7} \;,
\]
which is a reasonably good description of the result in Fig.\ts 2 for
large $\ell$. Furthermore, we can obtain the ratio $P_{\rm
B}/P_\kappa$ in the limit of large $\ell\gg \ell_{\rm m}$, which
yields
\[
{P_{\rm B}(\ell)\over P_\kappa(\ell)}
\approx \rund{\ell\over 4.9\times 10^5}^{1.2}\;,
\]
and roughly predicts the correct crossing point between these two
power spectra seen in Fig.\ts 2.

   \begin{figure} 
   \includegraphics[width=9cm]{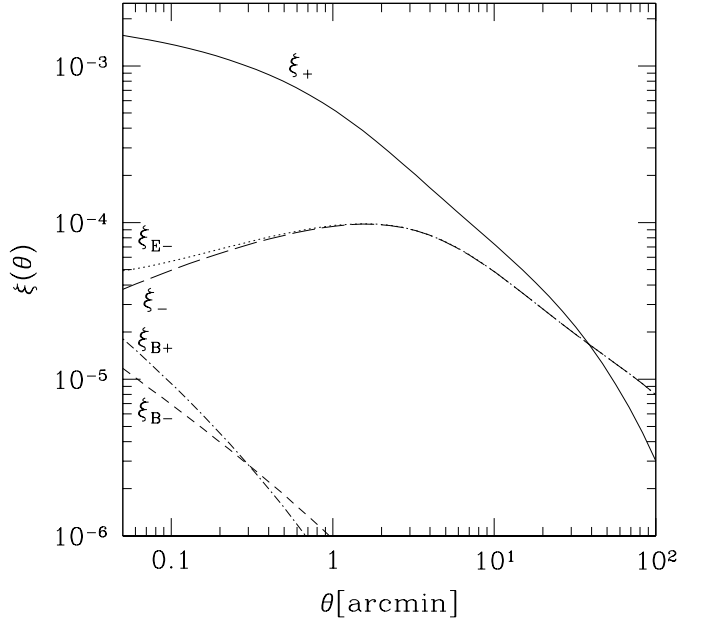}
   \caption{For the same model as in Fig.\ts 2, several correlation
   functions are plotted. The solid line shows $\xi_+(\theta)$; in
   fact, the correlation function $\xi_{\rm E+}$ cannot be
   distinguished from $\xi_+$ on the scale of this figure; their
   fractional difference is less than 1\%, even on the smallest scale
   shown. The two B-mode correlation functions are shown as well as
   $\xi_-$ and $\xi_{E-}$. Note that the difference between the latter
   two is larger than that of the corresponding `+'-correlation
   functions. } \end{figure}

In Fig.\ts 3 we have plotted several correlation functions; they have
been calculated from the power spectra plotted in Fig.\ts 2 by using
(\ref{eq:corrfcn}). The first point to note is that $\xi_{\rm E+}$
differs from $\xi_+$ by less than 1\% for angular scales larger than
$3''$; hence, the relative contribution caused by the source
correlation is even smaller than that seen in the power spectra. This
is due to the fact that the correlation function is a filtered version
of the power spectra, however with a very broad filter. This implies
that even at small $\theta$ scales, the correlation function is not
dominated by large values of $\ell$, where the contribution from source
clustering is largest, but low values of $\ell$ contribute
significantly. The influence of source clustering on the `$-$' modes
is larger, since the filtering function for those are narrower [i.e.,
${\rm J}_4(x)$ is a more localized function that ${\rm J}_0(x)$], and
$\xi_{\rm E-}$ differs from $\xi_-$ appreciably on scales below about
$1'$. 

Finally, in Fig.\ts 4 we have plotted the aperture measures. On scales
below about $1'$, the dispersion of $M_\perp$ is larger than about 1\% of
that of $M_{\rm ap}$. Hence, the ratio of these E- and B-mode aperture
measures are very similar to that of the corresponding power spectra.

   \begin{figure} 
   \includegraphics[width=9cm]{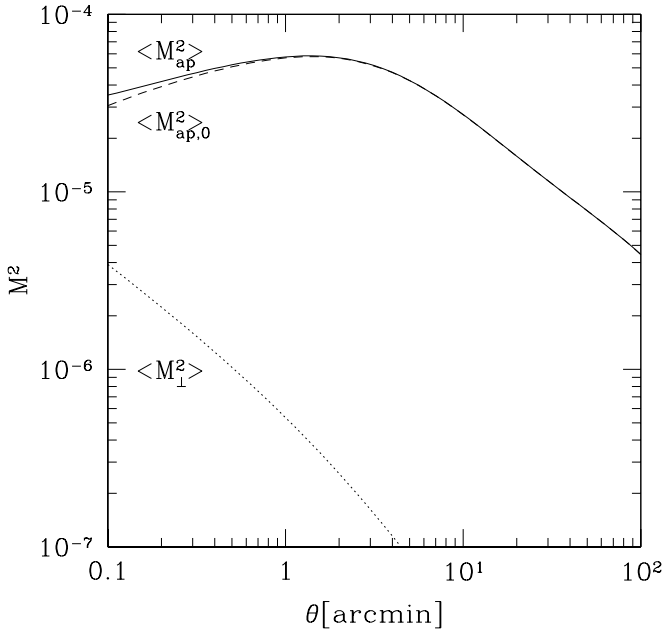}
   \caption{Aperture measures, for the same model as used in Fig.\ts
   2. Shown here is the dispersion of the aperture mass,
   $\ave{M_{\rm ap}^2}$, the corresponding function in the
   absence of source correlations (noted by the subscript `0') and
   ${\ave{M_\perp^2}}$, which is the aperture measure for the
   B-mode. As expected from the power spectra shown in Fig.\ts 2, and
   the fact that the aperture measures are a filtered version of the
   power spectra with a very narrow filter function, the B-mode
   aperture measure is considerably smaller than $M_{\rm ap}$ itself.}
   \end{figure}

The fact that $P_{\rm cE}$ and $P_{\rm B}$ are very similar in
amplitude means that by measuring $P_{\rm B}$, one can make an
approximate correction of $P_{\rm E}$, obtaining a value close to
$P_\kappa$ by subtracting $P_{\rm B}$ from $P_{\rm E}$. Owing to the
relative amplitude of these correlation-induced powers, such a
correction may be needed in future high-precision measurements of the
cosmic shear. 

\section{Discussion}
We have shown that the clustering of galaxies from which the shear is
measured leads to the presence of a B-mode in the cosmic shear field,
in addition to providing an additional component to the E-mode. The
reason for this effect in essence is the angular separation-dependent
redshift correlation of galaxies, which causes the mean of the product
of the angular-diameter distance ratio along two lines-of-sight not to
factorize, but to depend on $\theta$. For a fiducial model considered
in detail, the B-mode contribution amounts to more than $\sim 2\%$ on
angular scales below $1'$ (or $\ell \ga 2.16 \times 10^4$), and its
relative importance quickly rises towards smaller angles. On
substantially larger angular scales, however, the B-mode contribution
is small. Furthermore, the additional E-mode contribution is very
similar in size to the B-mode power, which will allow an approximate
correction of the measured E-mode for this additional term.

From an observational point-of-view, the most easily accessible
quantities are the shear correlation functions $\xi_\pm$, as one can
easily deal with gaps in the data field. In Sect.\ts 2 we have given
explicit relations regarding how other two-point statistics of the shear can be
calculated in terms of the shear correlation function. The finite
support of the functions $T_\pm$ indicates that the aperture measures
are more easily obtained from observational data than either the E-
and B-mode correlation functions, or the E- and B-mode shear
dispersions. Therefore, the aperture measures are the preferred method
to check for the presence of a B-mode contribution in the shear data.

We have varied some of the model parameters;
in particular, we have considered the case of lower mean source
redshift (corresponding to a brighter flux threshold), and
simultaneously increasing $A(1')$, such that the clustering length
$r_0$ stays about the same. In this case we found a very similar ratio
between the B- and E-mode power spectra as for the example considered
in Sect.\ts 4. We consider it unlikely that the observed B-mode in the
present day data sets is due
to the source clustering effect. The B-mode found in van
Waerbeke et al.\ts (2001) and Pen et al.\ (2002) can
actually be used to search quantitatively for residual systematics.
Its detection in van Waerbeke et al.\ts (2001)
was done by obtaining $M_{\rm ap}$ and $M_\perp$
by laying down a grid of circular apertures on the data field. A
more accurate measurement of $\ave{M_{\rm ap}^2}$ and
$\ave{M_{\perp}^2}$ has been obtained from the same data by Pen et
al.\ (2002), by calculating them from the observed correlation
functions $\xi_\pm$, as in (\ref{eq:N29}). In fact a subsequent analysis
revealed that the B-mode measured in these data were essentially residual
systematics caused by an overcorrection of the PSF, and can be
corrected for (van Waerbeke et al.\ 2002). In this latter analysis, no
significant B-modes are detected at small angular scales, but on
scales above $\sim 10'$, slightly significant values of $M_\perp$ are
detected; the effect discussed in this paper can certainly not account
for them.

The effect considered here seems to have been overlooked
hitherto. Bernardeau (1998) considered the effects of source
clustering on cosmic shear statistics and concluded that this source
clustering can strongly affect the skewness and kurtosis of the cosmic
shear, but to first order leaves the shear dispersion (and thus the
power spectrum) unaffected. Hamana et al.\ (2002) studied this effect
with ray tracing simulations, again concentrating on the skewness.
Most of the other ray tracing simulations of weak lensing (e.g., van
Waerbeke et al.\ 1999; Jain et al.\ 2000) assumed all sources to be at
the same redshift, in which case the additional power discussed here
does not occur. Lombardi et al.\ (2002) calculated the effect of
source clustering on the noise of weak lensing mass maps, showing that
it can provide a significant noise contribution in the inner regions
of clusters.

It must be pointed out that the effect considered here is unrelated to
other lensing effects which in principle could generate a B-mode, such
as lens-lens coupling or the break-down of the Born approximation (see
Bernardeau et al.\ 1997 and SvWJK for a discussion of
these two effects on the skewness). Numerical estimates (e.g., Jain et
al. 2000) show that these latter two effects are very weak. Bertin \&
Lombardi (2001) considered the situation of lensing by two mass
concentrations along the line-of-sight, where a B-mode is generated by
a strong lens-lens coupling, but the fraction of lines-of-sight where
this occurs is tiny. Another effect which could in principle generate a
B-mode from lensing is the fact that the observable is not the shear
itself, but the reduced shear (Schneider \& Seitz 1995). In the
appendix we show that this effect is indeed negligible.

Like the intrinsic alignment of galaxies, which can yield a spurious
contribution to the measured cosmic shear, the source clustering
effect can in principle be avoided if redshift estimates of the source
galaxies are available. In that case, by estimating the shear
correlation function, pairs of galaxies with a large likelihood to be
at the same distance can be neglected. In contrast to the intrinsic
correlation of galaxies, the B-mode from source clustering appears to
be fairly insensitive to the redshift distribution of the source
galaxies, provided the clustering length is kept fixed.

\appendix

\section{B-mode from reduced shear?}
The shear is not directly an observable, but is estimated from the
image ellipticities of distant galaxies. The expectation value of the
image ellipticity, however, is not the shear, but the reduced shear
$g=\gamma/(1-\kappa)$. Hence, the correlation of the observed
ellipticities is the correlation of the reduced shear, not the shear
itself. In cosmic shear, $\abs{\kappa}\ll 1$ nearly everywhere, and so
the difference between shear and reduced shear shall not play a big
role. However, at least a priori, this effect cannot be neglected, as
seen from the following argument:

The skewness $S_3=\ave{X^3}/\ave{X^2}^2$, where $X$ is a measure of
shear (such as $M_{\rm ap}$, or the reconstructed $\kappa$) has been
calculated by van Waerbeke et al.\ (2001) to be of order a few hundred.
On a scale of about one arcminute, $\ave{X^2}\sim 5\times 10^{-4}$, so
that $\ave{X^3}\sim 0.1 \ave{X^2}$, taking $S_3\sim 200$ for the
top-hat smoothed $\kappa$. The difference between the correlation
functions involving $g$ and those involving $\gamma$ is in principle
of the same order-of-magnitude as $\ave{X^3}$ and thus can be present
at the level of a few percent, and there is no reason why it should
not contain a B-mode contribution. 

We define the correlation functions
\be
\xi^g_\pm(\theta)=\ave{g_{\rm t}(\vc 0)g_{\rm t}(\vc \theta)}
\pm\ave{g_\times (\vc 0)g_{\times}(\vc \theta)}
\ee
and choose $\vc\theta=(\theta,0)$, so that $g_{\rm t}=-g_1$,
$g_\times=-g_2$. Using the approximation $g\approx\gamma(1+\kappa)$,
valid for $|\kappa|\ll 1$, we obtain
\be
\xi^g_\pm(\theta)=\xi_\pm(\theta)+\Delta\xi_\pm(\theta)\;,
\ee
where
\bea
\Delta\xi_\pm(\theta)&=&
\ave{\gamma_1(\vc 0)\gamma_1(\vc\theta)\eck{\kappa(\vc
0)+\kappa(\vc\theta)}}\nonumber \\
&\pm& \ave{\gamma_2(\vc 0)\gamma_2(\vc\theta)\eck{\kappa(\vc
0)+\kappa(\vc\theta)}}\;.
\eea
Replacing the shear and convergence by their Fourier transforms, this
becomes 
\bea
\Delta\xi_\pm(\theta)\!\!&=&\!\!\int{\d^2\ell_1\over (2\pi)^2}
\int{\d^2\ell_2\over (2\pi)^2}\int{\d^2\ell_3\over (2\pi)^2}
\ave{\hat\kappa(\vc\ell_1)\hat\kappa(\vc\ell_2)\hat\kappa(\vc\ell_3)}
\nonumber \\
&\times&\!\! \rund{ {\rm e}^{-{\rm i}\vc \ell_2\cdot \vc\theta}
+{\rm e}^{-{\rm i}(\vc \ell_2+\vc\ell_3)\cdot \vc\theta}}
\cos\eck{2\rund{\beta_1\mp\beta_2}}\;,
\elabel{69}
\eea
where, as before, $\beta_i$ is the polar angle of $\vc \ell_i$. The
triple correlator vanishes unless the sum of the wave-vectors equals
zero; one defines the bispectrum by
\be
\ave{\hat\kappa(\vc\ell_1)\hat\kappa(\vc\ell_2)\hat\kappa(\vc\ell_3)}
=(2\pi)^2\,\delta(\vc\ell_1+\vc\ell_2+\vc\ell_3)\,
b(\vc \ell_1,\vc\ell_2,\vc\ell_3)\;.
\ee
Performing the $\ell_3$-integration in (\ref{eq:69}) yields
\bea
\Delta\xi_\pm(\theta)\!\!&=&\!\!\int{\d^2\ell_1\over (2\pi)^2}
\int{\d^2\ell_2\over (2\pi)^2}
b(\vc\ell_1,\vc\ell_2,-\vc\ell_1-\vc\ell_2)\nonumber \\
&\times&\!\! \rund{ {\rm e}^{-{\rm i}\vc \ell_2\cdot \vc\theta}
+{\rm e}^{{\rm i}\vc \ell_1\cdot \vc\theta}}
\cos\eck{2\rund{\beta_1\mp\beta_2}}\;.
\elabel{71}
\eea
The function $b(\vc\ell_1,\vc\ell_2,-\vc\ell_1-\vc\ell_2)$ has three
independent arguments, namely the moduli $\ell_1$ and $\ell_2$, and
the angle $\phi=\beta_1-\beta_2$ between the two $\vc\ell$-vectors. We
therefore write $b(\vc\ell_1,\vc\ell_2,-\vc\ell_1-\vc\ell_2)
=\tilde b(\ell_1,\ell_2,\phi)$, make use of the symmetry in the
integrand of (\ref{eq:71}), and replace the $\beta_1$-integration by
one over $\phi$:
\bea
\Delta\xi_+(\theta)&=&2\int_0^\infty{\d\ell_1\,\ell_1\over (2\pi)^2}
\int_0^\infty{\d\ell_2\,\ell_2\over (2\pi)^2}
\int_0^{2\pi}\d\phi\;\tilde b(\ell_1,\ell_2,\phi)\nonumber \\
&\times &
\int_0^{2\pi}\d\beta_2\;
{\rm e}^{-{\rm i}\ell_2\theta\cos\beta_2}\,\cos(2\phi)
\nonumber \\
&=&{1\over \pi}
\int_0^\infty{\d\ell_1\,\ell_1\over (2\pi)}\,{\rm J}_0(\ell_1\theta)
\int_0^\infty{\d\ell_2\,\ell_2\over (2\pi)}
\nonumber \\
&\times &
\int_0^{2\pi}\d\phi\;\cos(2\phi)\,\tilde b(\ell_1,\ell_2,\phi)\;;
\eea
analogously, one obtains
\bea
\Delta\xi_-(\theta)&=&
{1\over \pi}
\int_0^\infty{\d\ell_1\,\ell_1\over (2\pi)}\,{\rm J}_4(\ell_1\theta)
\int_0^\infty{\d\ell_2\,\ell_2\over (2\pi)}
\nonumber \\
&\times&
\int_0^{2\pi}\d\phi\;\cos(2\phi)\,\tilde b(\ell_1,\ell_2,\phi)\;.
\eea
Inserting these expressions into (\ref{eq:N15}) and making use of 
(\ref{eq:N14}), one immediately sees that the reduced shear does not
yield any B-mode contribution, and that the correlation functions for
the reduced shear are
\be
\xi^g_\pm(\theta)=\int{\d\ell\,\ell\over 2\pi}
{\rm J}_{0,4}(\ell\theta)\eck{P_\kappa(\ell)+P^{(3)}(\ell)}\;,
\ee
where
\be
P^{(3)}(\ell)\!\!=\!\!2\int\!\!{\d^2\ell'\over(2\pi)^2}\,
\eck{{2(\vc\ell\cdot\vc\ell')^2\over
|\vc\ell|^2\,|\vc\ell'|^2}-1}\,
b(\vc\ell,\vc\ell',-\vc\ell-\vc\ell') \,.
\ee
Thus, considering the reduced shear yields an additional E-mode power
to the one obtained from considering the shear itself.

\begin{acknowledgements}
We thank L.J. King for useful comments on the manuscript.
This work was supported by the TMR Network ``Gravitational Lensing:
New Constraints on Cosmology and the Distribution of Dark Matter'' of
the EC under contract No. ERBFMRX-CT97-0172 and by the German Ministry
for Science and Education (BMBF) through the DLR under the project 50
OR 0106.      

\end{acknowledgements}

\end{document}